\documentclass[aps,prb,twocolumn,reprint,amsmath,amssymb,superscriptaddress]{revtex4-1}
\usepackage{amsfonts}
\usepackage{mathrsfs}
\usepackage{amsmath}
\usepackage{color}
\usepackage{natbib}
\usepackage{graphicx}
\usepackage{bm}
\usepackage{amssymb}
\usepackage{xspace}
\usepackage{epstopdf}
\usepackage{dcolumn}
\usepackage{multirow}
\usepackage[colorlinks=true, letterpaper=true, pdfstartview=FitV, linkcolor=blue, citecolor=blue, urlcolor=blue]{hyperref}

\makeatletter

\newcommand{\Rmnum}[1]{\expandafter\@slowromancap\romannumeral #1@}
\makeatother

\begin{document}

\title{From Type-II Triply Degenerate Nodal Points and Three-Band Nodal Rings to Type-II Dirac Points in Centrosymmetric Zirconium Oxide}

\author{Ting-Ting Zhang}

\address{Institute of Physics, Chinese Academy of Sciences, Beijing 100190, China}
\address{School of Physical Sciences, University of Chinese Academy of Sciences, Beijing 100190, China}

\author{Zhi-Ming Yu}
\email{zhiming\_yu@sutd.edu.sg}
\address{Research Laboratory for Quantum Materials, Singapore University of Technology and Design, Singapore 487372, Singapore}

\author{Wei Guo}
\address{School of Physics, Beijing Institute of Technology, Beijing 100081, China}

\author{Dongxia Shi}
\address{Institute of Physics, Chinese Academy of Sciences, Beijing 100190, China}
\address{School of Physical Sciences, University of Chinese Academy of Sciences, Beijing 100190, China}
\address{Beijing Key Laboratory for Nanomaterials and Nanodevices, Beijing 100190, China}

\author{Guangyu Zhang}
\email{gyzhang@aphy.iphy.ac.cn}
\address{Institute of Physics, Chinese Academy of Sciences, Beijing 100190, China}
\address{School of Physical Sciences, University of Chinese Academy of Sciences, Beijing 100190, China}
\address{Collaborative Innovation Center of Quantum Matter, Beijing 100190, China}
\address{Beijing Key Laboratory for Nanomaterials and Nanodevices, Beijing 100190, China}

\author{Yugui Yao}
\email{ygyao@bit.edu.cn}
\address{School of Physics, Beijing Institute of Technology, Beijing 100081, China}

\begin{abstract}
Using first-principles calculations, we report that ZrO is a topological material with the coexistence of three pairs of type-II triply degenerate nodal points (TNPs) and three nodal rings (NRs), when spin-orbit coupling (SOC) is ignored. Noticeably, the TNPs reside around Fermi energy with large linear energy range along tilt direction ($> 1  \ \rm{eV}$) and the NRs are formed by three strongly entangled bands. Under symmetry-preserving strain, each NR would evolve into four droplet-shaped NRs before fading away, producing distinct evolution compared with that in usual two-band NR. When SOC is included, TNPs would transform into type-II Dirac points while all the NRs are gaped. Remarkably, the type-II Dirac points inherit the advantages of TNPs: residing around Fermi energy and exhibiting  large linear energy range. Both features  facilitate the observation of interesting phenomena induced by type-II dispersion. The symmetry protections and low-energy Hamiltonian for the nontrivial band crossings are discussed.
\end{abstract}
\maketitle

\section{Introduction }

The exploration of topological materials has now extended from topological insulators \cite{QiRMP,HasanRMP} to topological semimetals, such as Weyl \cite{2011Wan,HgCrSe2011,2015PRXTaAs} and Dirac semimetals \cite{2012Na3Bi,2013Cd3As2,Sheng2014,WangPSS2017}.
In Weyl (Dirac) semimetals, the low energy electrons residing around Weyl (Dirac) point obey relativistic equation, providing  the possibility to simulate  intriguing high-energy physics in solids \cite{YangSpin,Guan2017}. Currently, the
discovery of type-II Weyl semimetals \cite{Soluyanov2015,Zhou_2016Natphy_MoTe2} unveils that in solids, the kinds
of quasiparticles would be more abundant than that in high-energy
physics, due to the reduced symmetry constraint. Soon after, type-II
Dirac point and multi-fold (beyond doubly and fourthly) degenerate nodal
point are successively proposed \cite{Chang2017,2017YPd2Sn,Duan2016,Science_multi,Weng_TaN,Weng_ZrTe,TaS2017,2017VAl3,ZhuPRX,CongCaAgBi}.
All the new fermions exhibit physical  phenomena distinguished from each other and conventional Weyl and Dirac fermions \cite{Yu2016,Beenakker2016,2017VAl3,ZhuPRX}. Many materials are predicted to be the candidates
for hosting the new fermions and some of them are confirmed by experiment, such as, ${\rm MoTe_{2}}$ \cite{Zhou_2016Natphy_MoTe2,MoTe2_2016PRX} (${\rm PtTe_{2}}$ \cite{exp_Park2017PdTe2,2017zhou_PtTe2,Zhou_PtSe2,PdTe2_2017PRB}) families are confirmed as
type-II Weyl   (Dirac) semimetal and $\rm{MoP}$ \cite{2017NatureMoP} is confirmed
as type-I TNP semimetal.
However, searching for ideal semimetals \cite{ZhangPRL2016} with nodal point locating at Fermi energy and exhibiting large linear energy
range is still desirable.

Besides the topological semimetals with zero dimensional band crossing, NR semimetals, featuring one-dimensional band crossing \cite{Weng2015Gnetwork,Cu3PdN2015PRL,Ca3P22016,Yang_type2,Chen2016PRL,Sheng_JPCL,HasanPbTaSe2,ZrSiS,Yang_hopflink,ZhangJPCL2017,ChangPRL2017}, also have attracted tremendous attention. Intuitively, similar to Weyl semimetal, a two band model is enough to capture the low-energy physics of (doubly degenerate) NR. The two-band model works well for most reported NR materials.
However, a  recent work shows that  the NR in ${\rm TiB_{2}}$ \cite{Zhang2017} is distinct from conventional two-band NR in that its formation  requires a four-band model, as  the two bands forming  NR are strongly entangled with other two bands.
Hence,  the NR in ${\rm TiB_{2}}$   are termed as four-band NR \cite{Zhang2017}.
Furthermore, the evolution of two-band NR and four-band NR under  strain can be very different \cite{Fangchen2015PRB,Zhang2017}.
Consequently, one can expect that  other multi-band NR may lead to many distinct phenomena compared with  conventional two-band NR and may be  considered as a new kind of intriguing topological semimetal to be discovered.

In this work, using  first-principles calculations and symmetry analysis, we show that ${\rm ZrO}$ is a topological metal with the coexistence of three pairs of type-II TNPs and  three NRs, when SOC is ignored.
The type-II TNPs locate at three high-symmetry lines and are protected by the $C_{4v}$ point group symmetry.
TNPs have been  predicted  in many materials \cite{Zhang2017,Weng_ZrTe,Weng_TaN,TaS2017,ZhuPRX}.
However, type-II TNP semimetal is rare\cite{Chang2017}. Compared with previously studied TNP semimetals, the  TNPs identified here has several advantages, such as it has type-II dispersion, its nodal energy is close to Fermi energy and its linear energy range is large.
Regarding to the three NRs, we find that they lie in three mirror planes, respectively and  hence are doubly protected by mirror symmetry and the combination of time reversal  ($\cal{T}$) and inversion ($\cal{P}$) symmetry.
Interestingly, the formation of the NRs requires three bands rather than two bands and hence we term the NRs here as three-band nodal ring (TNR).
Moreover, it is found the three bands forming TNRs are exactly the bands forming TNPs.
In the following, we will see that the appearance of TNRs is closely related to the TNPs and the coexistence of TNR and TNP can be found in many materials \cite{Zhang2017,ZhuPRX}.
Particularly, under symmetry-preserving strain, e.g. hydrostatic strain, each TNR here  would evolve into four droplet-shaped nodal rings before vanishing, distinguished from previously studied nodal rings.
When SOC is included, all the TNRs are gapped while each type-II TNP becomes a type-II Dirac point due to the presence of $\cal{P}$ and $\cal{T}$ symmetry.
Remarkably, all the type-II Dirac points  almost locate at Fermi energy  with large linear energy range.
Thus, ${\rm ZrO}$ would be an ideal platform for studying the interesting physics induced by type-II dispersion.

\section{Method and crystal structure }
We have employed the Vienna ab initio simulation package (VASP)  \cite{VASP1996} for
most of the first-principle calculations.  The potentials are treated with the projector augmented wave (PAW) method \cite{PAW1994}. Exchange-correlation potential
is treated within the generalized gradient approximation (GGA) of
the Perdew-Burke-Ernzerhof type \cite{PBE1992}. The cutoff energy is chosen as $520\ {\rm eV}$
and a $21\times21\times21$ $\Gamma$-centered
$k$-mesh for self-consistent calculations. The energy convergence
criterion is set to be $10^{-6}\ {\rm eV}$. The crystal structure
is fully relaxed until the maximum force on each atom was less than
$0.01\ {\rm eV}/ \rm{\AA}$.

\begin{figure}[t]
\includegraphics[width=8cm]{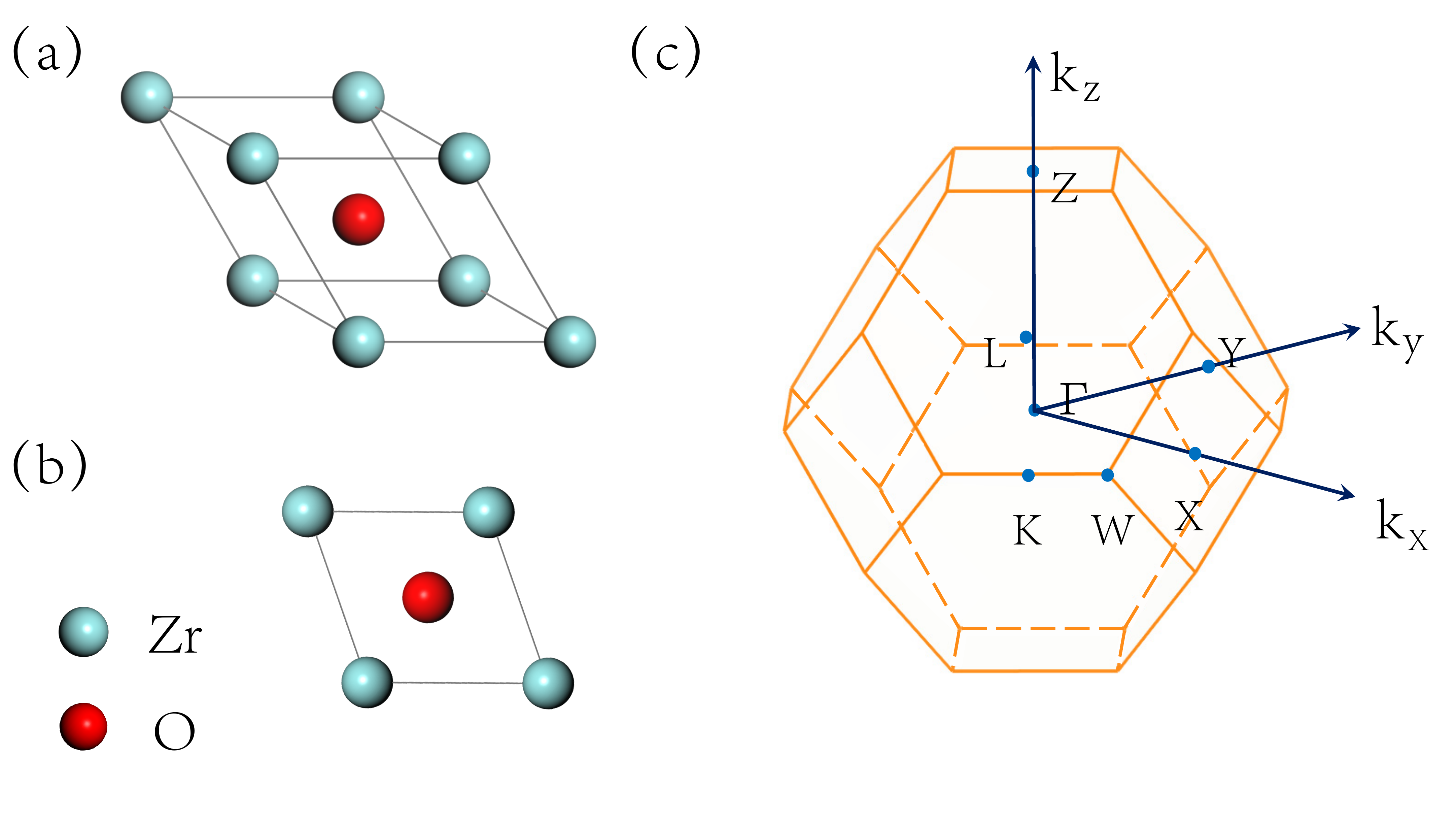}
\caption{ (a) Side view and (b) top view of the crystal structure of FCC-type $\rm{ZrO}$.
(c) The bulk Brillouin zone of $\rm{ZrO}$. \label{fig1}}
\end{figure}

The elements ${\rm Zr}$ and ${\rm O}$ can form different  kinds of zirconium
oxide \cite{ZrTaNO1954,1981ZrO}. ${\rm ZrO}$ can be synthesized
at proper oxygen atmosphere and shows FCC-type cubic crystal structure
with space group $Fm\bar{3}m$ (No. 225, $O_{h}^{5}$) as shown in
Fig. \ref{fig1}. ${\rm Zr}$ and ${\rm O}$
atoms occupy the $(0,0,0)$ and $(0.5,0.5,0.5)$ Wyckoff positions,
respectively.
The experimental lattice constants are  $a=b=c=3.254\ \rm{\AA}$. The optimized lattice constants are $a=b=c=3.271\ \rm{\AA}$, which are very closed to experimental values, being overestimated by about 0.5\% and all the results discussed in
the following are from the calculations with optimized structures.
Figure \ref{fig1}(c) shows the bulk Brillouin zone (BZ) of ${\rm ZrO}$
crystal.

\section{Triply degenerate nodal point}

\begin{figure}[t]
\includegraphics[width=8.5cm]{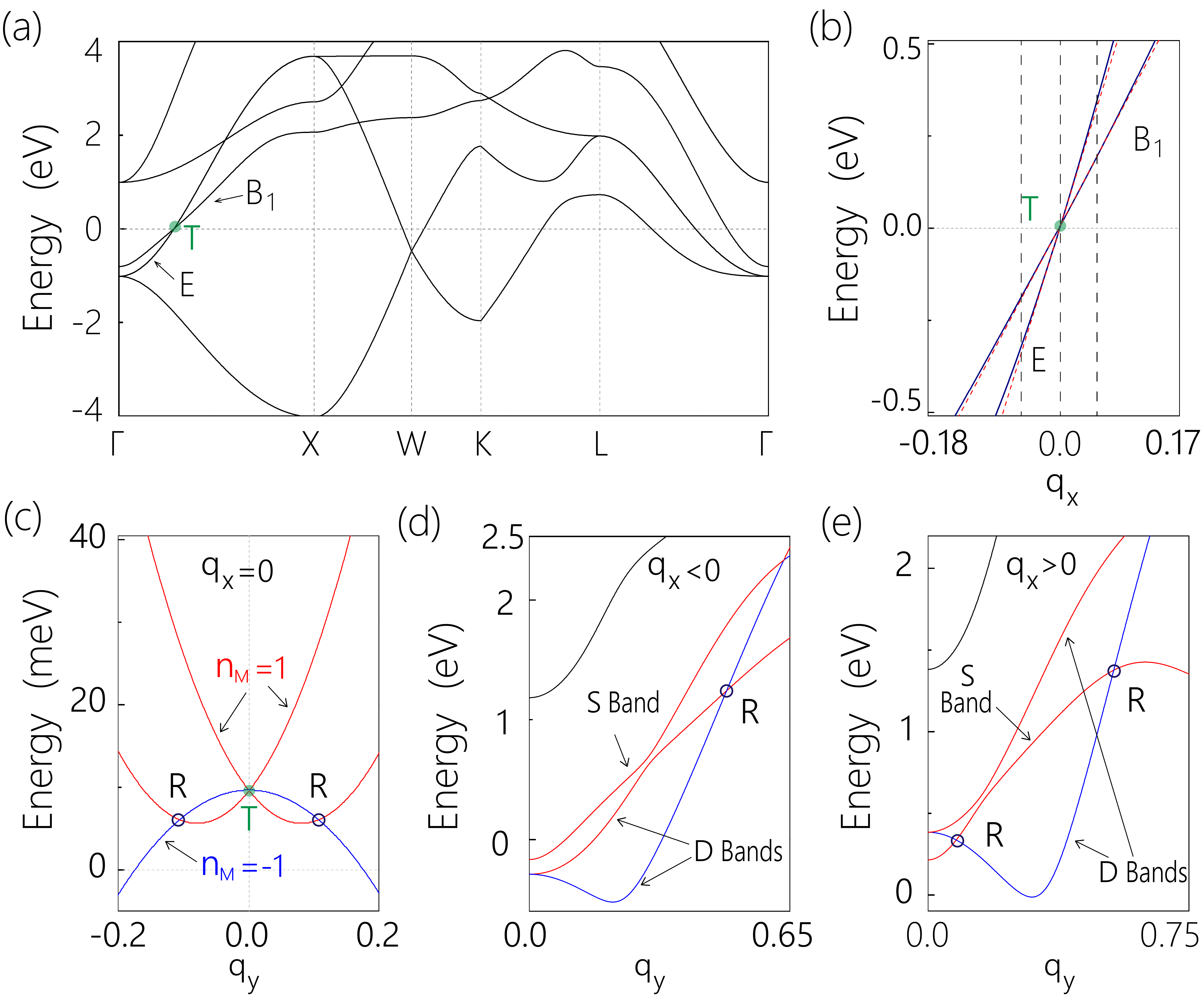}
\caption{(a) Calculated electronic band structure of ${\rm ZrO}$ without SOC.
The crossing point on $\Gamma$-$X$ line is triply degenerate, labeled
as $T$. The crossing bands have different IRs of $C_{4v}$ point
group: $E$ and $B_{1}$. (b) Dispersion around the TNP ($T$ point)
by first-principles calculations (solid lines) and $k\cdot p$ model
fitting (red dashed lines). (c), (d) and (e) are the band dispersion along $q_y$ direction
in the $q_{z}=0$ plane for different $q_{x}$ marked by the three black dashed lines in (b). The two  mirror eigenvalues of  ${\cal{M}}_{xy}$ ($n_{M}=1$ and $n_{M}=-1$) of the relevant three bands are denoted by red and
blue color, respectively. The crossing points belonging to TNR are
labeled as $R$. Here, the unit of $q_{x(y)}$ is $\pi/a$ ($\pi/b$).
\label{fig2}}
\end{figure}

The electronic band structure of ${\rm ZrO}$ without SOC is ploted in Fig.\ref{fig2}(a), showing a metallic phase with highly dispersive bands.
The band crossing point $T$ on $\Gamma$-$X$ path is especially striking, as it is close to Fermi
energy and features large linear energy range ($> 1\ \rm{eV}$). On $\Gamma$-$X$ path, the litter point group is $C_{4v}$.
And the two bands formed point $T$ belong to two distinct irreducible representations (IRs) of $C_{4v}$:
$B_{1}$ (one-dimensional IR) and $E$ (two-dimensional IR), respectively (see Tab.\ref{tab1}).
Therefore, the band crossing is triply degenerate and is protected by $C_{4v}$ symmetry. Moreover, the slope of the two crossing bands share same sign. Thus, point $T$ is a type-II TNP.
Due to the three $C_{4v}$  symmetries along $k_{x}$, $k_{y}$ and $k_{z}$ axes, ${\rm ZrO}$ exhibits three pairs of type-II TNP residing on three axes.
In the following, we use the point $T$ to discuss the properties of TNP.
Besides, the  band crossing at W point is also obvious, which is about 0.47 eV below Fermi energy. This band crossing  is doubly degenerate and is essential, as its IR is  $E$ of the litter group of W point $D_{2d}$.

To characterize the low-energy physics of type-II TNPs, we establish a $k\cdot p$  model in the vicinity of point $T$ using
bands with $B_{1}$ and $E$ IRs as basis.
The Hamiltonian around $T$ up to linear order in $\bm{q}$ (measured from $T$ point) reads (see Supporting Information for details)
\begin{eqnarray}
{\cal H}_{T} & = & C_{1}q_{x}+\left(\begin{array}{ccc}
C_{2}q_{x} & Dq_{z} & -Dq_{y}\\
Dq_{z} & -C_{2}q_{x} & 0\\
-Dq_{y} & 0 & -C_{2}q_{x}
\end{array}\right),
\label{THam}
\end{eqnarray}
where the model parameters $C_{1(2)}$, $D$ are real and the first
term in ${\cal H}_{T}$ denotes the tilt effect. Along $q_{x}$  axis ($\Gamma$-$X$ path),
the dispersion of model ${\cal H}_{T}$ is $\varepsilon=(C_{1}\pm C_{2})q_{x}$ indicating a triply degenerate point at $q_{x}=0$.
Moreover, by fitting of this model to the DFT band structure {[}see Fig. \ref{fig2}(b){]},
 we find $C_1=6.607$ eV$\cdot$\AA, $C_2=3.827$ eV$\cdot$\AA ~and $D=0.114$ eV$\cdot$\AA. Hence, one has $|C_{1}|>|C_{2}|$ which is the typical  feature of type-II nodal point.
Figure \ref{fig2}(b) also shows  the linear fitting works well at least up to $\sim1{\rm eV}$ energy range.
In Fig. \ref{fig2}(c), we  plot the dispersion of TNP along the $q_{y}$ direction in the $q_z=0$ plane.
In contrast, the dispersion along $q_{y}$ direction does not have energy tilt, due to the presence of  mirror symmetry (${\cal M}_{xz}$) with respect to $x$-$z$ plane.
The dispersion along $q_{z}$ direction is identical with that along $q_{y}$ direction as guaranteed by  the $C_{4x}$ rotation symmetry with respect to $\Gamma$-$X$ path.
The $C_{4x}$ rotation symmetry also manifests itself in Hamiltonian (\ref{THam}), e.g. the  coefficients of $q_y$ and $q_z$ in ${\cal{H}}_T$ are the same (up to their sign).

\begin{table}[t]
\begin{tabular}{|c|c|c|c|c|c|}
\hline
$\ \ \ C_{4v}\ \ \ $ & $\ \ \ E\ \ \ $ & $\ \ \ 2C_{4}\ \ \ $ & $\ \ \ C_{2}\ \ \ $ & $\ \ \ 2\sigma_{v}({\cal M}_{xy})\ \ $ & $\ \ 2\sigma_{d}\ \ $\tabularnewline
\hline
\hline
$B_{1}$ & 1 & -1 & 1 & 1 & -1\tabularnewline
\hline
$E$ & 2 & 0 & -2 & 0 & 0\tabularnewline
\hline
\end{tabular}\caption{Character table for the the two encountered representations of $C_{4v}$ point group on $\Gamma$-$X$ path.\label{tab1}}
\end{table}

\section{Three-band nodal ring}

From Fig. \ref{fig2}(c), one observes that near TNP ($T$ point) there exist two doubly degenerate points ($R$), protected by ${\cal M}_{xy}$ symmetry as we will discuss later.
Due to the presence of  $\cal{PT}$  symmetry in ${\rm ZrO}$, point $R$ would not exist in isolation but hint the appearance of NR \cite{Weng2015Gnetwork}.
In the  following, we  discuss the NR containing point $R$.

\begin{figure*}
\includegraphics[width=13cm]{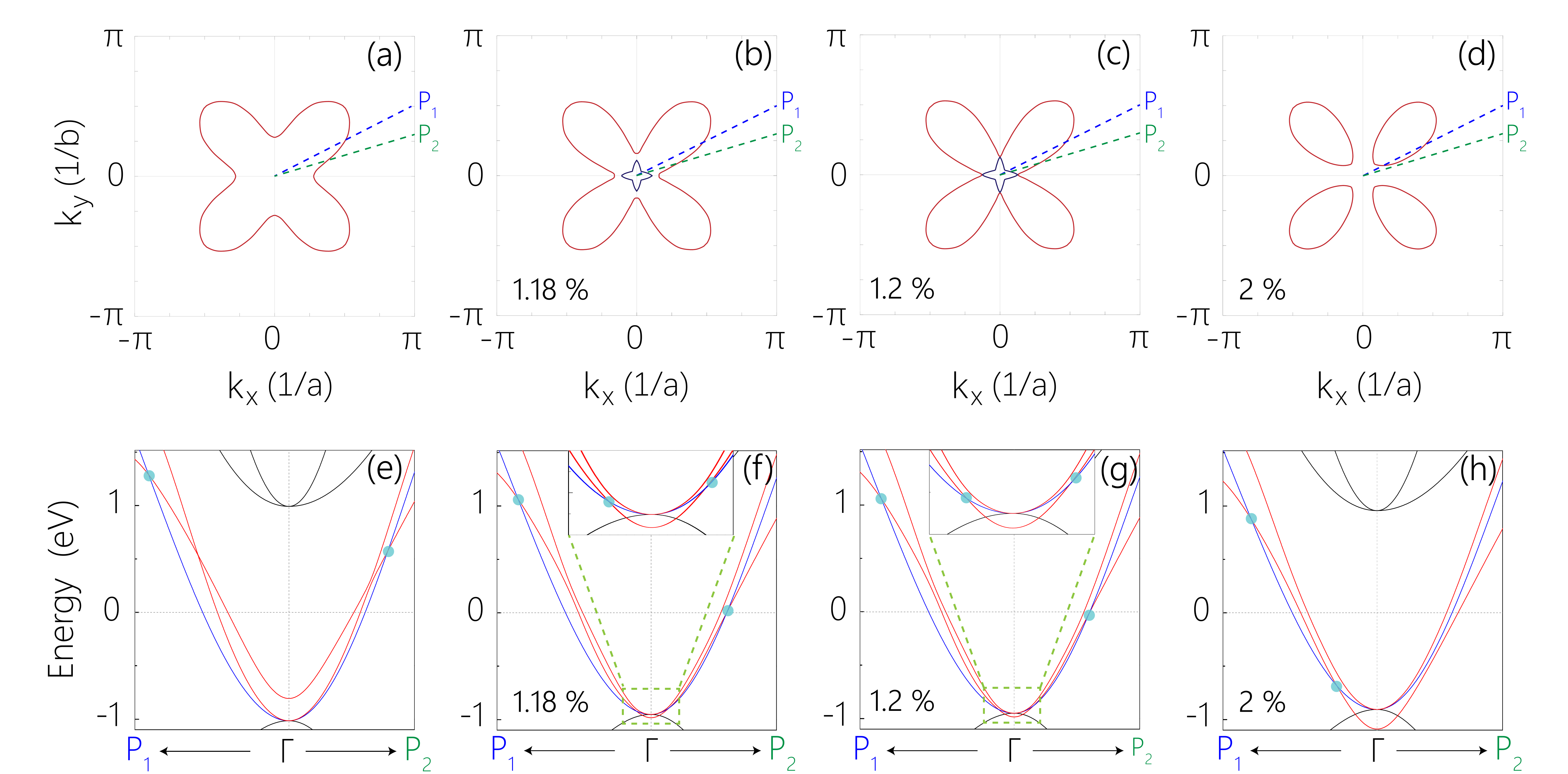}
\caption{The shape of  TNR of $\rm{ZrO}$ under the hydrostatic strain of (a) $0\%$, (b) $1.18\%$, (c) $1.2\%$ and (d) $2\%$.
(e-h) Electronic band structures of $\rm{ZrO}$ along two paths denoted by blue ($\Gamma$-$P_1$) and green ($\Gamma$-$P_2$) dashed lines in (a-d) for different strain.
In (e-h), the red  lines and blue line are the three bands forming TNR with mirror eigenvalues $n_M=1$ and $n_M=-1$, respectively. The band crossing points are labeled by the cyan dots.  \label{fig3}}
\end{figure*}

First, we present the band dispersion along $q_{y}$ direction for   a constant $q_x$ value setting
below ($q_x<0$) and  above ($q_x>0$) the TNP {[}Fig. \ref{fig2}(d) and \ref{fig2}(e){]}.
In these two cases, there also exist  band crossing points (also labeled as $R$), indicating the NR may lie in $k_{z}=0$ plane.
Indeed, a careful scan  of  band dispersion  in $k_{z}=0$ plane  shows there exists  a NR centered at $\Gamma$ point with four-leaf clover-like shape, as shown in Fig. \ref{fig3}(a).
Particularly, from Fig. \ref{fig2}(c), one finds the formation of the NR here requires three bands. Hence  we term it as TNR.
TNR has never been  studied in previous works.

Then, we preform a symmetry analysis on the formation of TNR.
For simplification, we divide the three relevant bands as $D$ bands and $S$ band, as indicated in Fig. \ref{fig2}(d) and \ref{fig2}(e).
$D$ bands are the two bands which would stick together on $\Gamma$-$X$ line {[}containing the $q_{y}=0$ point in Fig. \ref{fig2}(c)-\ref{fig2}(e){]} and $S$ band  is the remaining one.
With symmetry analysis, one knows:
(i), $D$ bands, which are doubly degenerate at $q_{y}=0$ point {[}Fig. \ref{fig2}(d) and \ref{fig2}(e){]}, would split into two nondegenerate bands at generic momentum  points in $q_{z}=0$ plane, as such points do not have $C_{4v}$ symmetry.
(ii), $D$ and $S$ bands are in $q_{z}=0$ mirror plane (containing $\Gamma$-$X$ line) and hence have explicit  eigenvalues of mirror symmetry ${\cal M}_{xy}$, given as  $n_{M}=\pm1$.
Also, we know the IRs  of $S$ and $D$ bands on $\Gamma$-$X$ line are $B_1$ and  $E$ [see Fig. \ref{fig2}(a)], and the character for ${\cal M}_{xy}$ symmetry  of  $B_1$ and  $E$ IRs are $1$ and $0$ (see Tab. \ref{tab1}).
Thus, the   ${\cal M}_{xy}$ eigenvalues  of $S$ band   and  $D$ bands are $n_M=1$ and $n_M=\pm1$, respectively.
As a result, band crossing can happen between  the three bands  with different $n_M$  and the degeneracy of the crossing points is two-fold, not   three-fold as that on $\Gamma$-$X$ line.
By analyzing the mirror eigenvalue of $S$ and $D$ bands in detail, it is found the upper (lower) branch of
$D$ bands has $n_{M}=1$ ($n_{M}=-1$) while the $S$ band has $n_{M}=1$
{[}see Fig. \ref{fig2}(d) and \ref{fig2}(e){]}, consistent with the above discussion.

For the case of $q_{x}<0$ (below TNP), $S$ band is higher than $D$ bands at $q_{y}=0$ point {[}Fig. \ref{fig2}(d){]}.
Since $S$ band and $D$ upper band share same mirror eigenvalue ($n_{M}=1$), they would feature anticrossing.
Meanwhile, the  $D$ bands  form  a band crossing between themselves as they have opposite $n_{M}$.
In sharp contrast, for $q_{x}>0$ (above TNP) , $S$ band is lower than $D$ bands at $q_{y}=0$ point {[}Fig. \ref{fig2}(e){]}, then  $S$ band can linearly cross with the lower band of $D$.
Consequently, the nodal points are formed by the  $D$ bands when $q_{x}<0$  and by the $S$ band and the lower band of $D$ when $q_{x}>0$, indicating that  the  band order inverse  between $S$ and $D$  bands at $q_x=0$ point  ($\Gamma$-$X$ line) is crucial for the formation of TNR.
Observe that this band inversion is inevitable due to the appearance of TNP (see Fig. \ref{fig2}a).
Thus, for TNP with   dispersion being similar with that in  Fig. \ref{fig2}(c), one can expect TNR may coexist with TNP.
Indeed, TNP with such  dispersion can be found in  many identified TNP materials \cite{Zhang2017,ZhuPRX} and lots of  experimentally synthesized  ZrO family materials (see Figure S1 in Supporting Information).

Since the evolution of two-band and four-band NR under strain are very different\cite{Fangchen2015PRB},
TNR proposed here under strain  may also show unique behavior.
Figure \ref{fig3}(a-d) show the shape of TNR  for four representative hydrostatic strains of $0. \%$, $1.18 \%$, $1.2 \%$ and  $2 \%$.
Figure \ref{fig3}(e-h) give the band dispersion of $\Gamma$-$P_{1(2)}$ line  under the corresponding  strains.
In Fig. \ref{fig3}(e-h), the relevant three bands are highlighted by red or blue, according to their ${\cal{M}}_{xy}$ mirror eigenvalues ($n_M=\pm1$). Here, we only focus on the band crossings between the three bands, e.g. cyan dots in Fig. \ref{fig3}(e-h).
Under positive strain, the order of bands at $\Gamma$ point would inverse [see Fig.  \ref{fig3}(e) and \ref{fig3}(f)].
This band inversion gives rise to a new nodal ring [see Fig. \ref{fig3}(b)], which can be clearly find from Fig.  \ref{fig3}(e-f) as  the number of the crossing point  on $\Gamma$-$P_{1(2)}$ line has changed from one [Fig.  \ref{fig3}(e)] to two [Fig.  \ref{fig3}(f)].
Increasing strain to a critical value  ($1.2 \%$), the new nodal ring would grow in size and finally  touch with the original one as shown in Fig.  \ref{fig3}(c).
Though the two nodal ring are touched together here, they are not nodal chain \cite{IrF4,wang2017hourglass,HfC2017PRL} as the two nodal rings are in the same plane and the touching  is  accident.
Remarkably, beyond the critical strain, the two nodal rings would  merge together to produce  four droplet-shaped nodal rings [see Fig.  \ref{fig3}(d)].
Keep increasing strain, the droplet-shaped nodal rings would fade away (not shown).
This unique evolution of TNR in ${\rm ZrO}$ results from the strong entanglement of the three bands, which can not be found in two-band NR.

\section{Type-II Dirac points}

\begin{figure}[t]
\includegraphics[width=8.5cm]{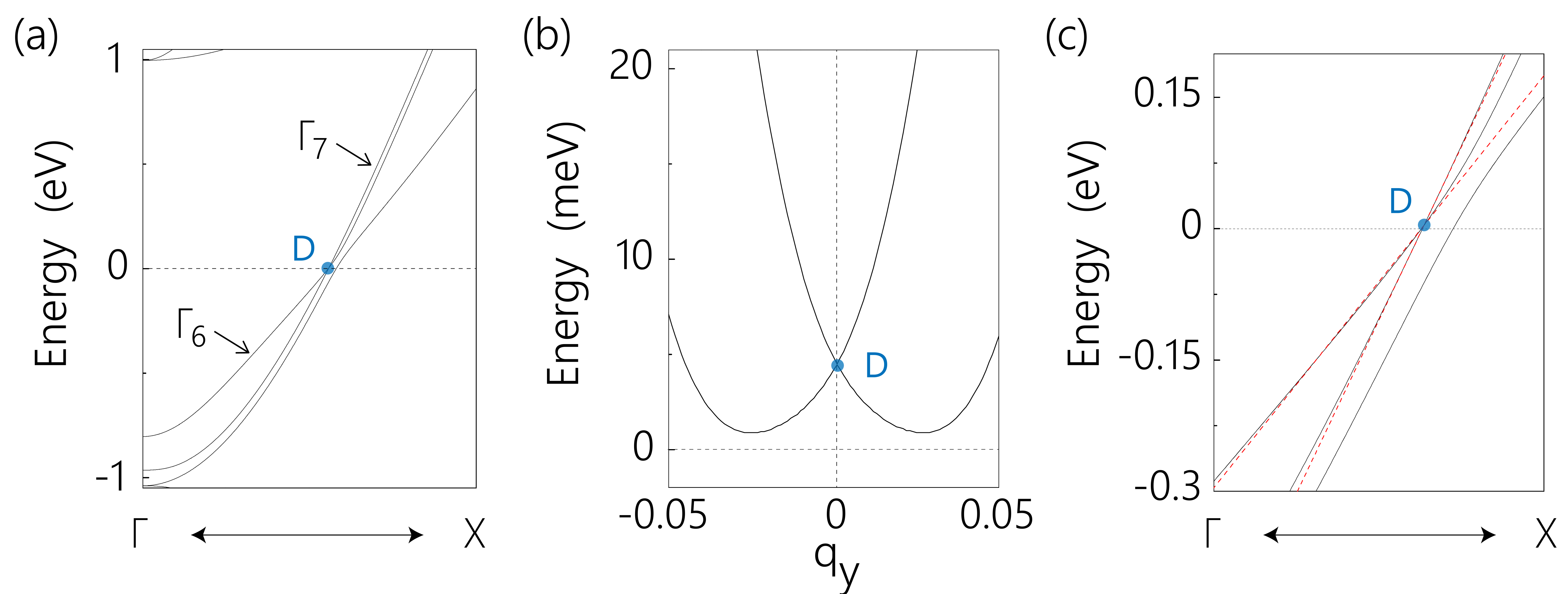}
\caption{(a) Band structure on the $\Gamma$-$X$ line with SOC included. The irreducible representation of bands forming Dirac point are indicated.
(b) The band dispersion of Dirac point along the $q_y$-direction. (c)
Enlargement of electronic band structure   around  Dirac point by first-principles calculations (solid lines) and $k\cdot p$
model fitting (red dashed lines).
The Dirac point is labeled as D.\label{fig4}}
\end{figure}

At last, we discuss the band dispersion of ${\rm ZrO}$ with SOC, which is plotted in  Fig. \ref{fig4}.  Due to the presence of ${\cal{P}}$ and ${\cal{T}}$ symmetry, each band in Fig. \ref{fig4} is at least doubly degenerate.
On $\Gamma$-$X$ path, the type-II TNP transforms into a type-II Dirac point formed by the bands with $\Gamma_{6}$ and $\Gamma_{7}$ IRs of $C_{4v}$ double group [Fig. \ref{fig4}(a)].
Also, the dispersion of Dirac node along $q_{y}$ direction in $q_{z}=0$ plane is shown in Fig. \ref{fig4}(b).
Due to the $C_{4x}$ rotation symmetry, the dispersion along $q_{z}$ direction is same with that along $q_{y}$ direction.
Using the $\Gamma_{6}$ and $\Gamma_{7}$ states at Dirac point as basis, the low-energy $k\cdot p$ Hamiltonian around Dirac point (up to linear order measured from Dirac point) can be established as
\begin{eqnarray}
{\cal H}_{D} & = & \left(\begin{array}{cc}
h_{+} & 0\\
0 & h_{-}
\end{array}\right),\label{DHam}
\end{eqnarray}
where each entry is a $2\times2$ matrix with
\begin{equation}
h_{\pm}=wq_{x}+v_{1}(q_{z}\sigma_{x}\pm q_{y}\sigma_{y})+v_{2}q_{x}\sigma_{z}.\label{WHam}
\end{equation}
Here, $\bm{\sigma}$ are Pauli matrix. The expression of $h_{\pm}$ denotes
a tilted Weyl Hamiltonian with $\pm1$ chirality, directly showing
this band crossing is a Dirac node. Coefficients $v_{1(2)}$ and $w$ represent  Fermi velocity and energy tilt, respectively.
By fitting Hamiltonian ${\cal H}_{D}$ to DFT results [see Fig. \ref{fig4}(c)], we find  $w=6.607$ eV$\cdot$\AA, $v_1=0.343$ eV$\cdot$\AA~and $v_2=3.925$ eV$\cdot$\AA, and have $|w|>|v_{2}|$ which is consistent with the type-II dispersion.
Meanwhile, the TNR and the band crossing at W point are gapped (see Figure S2 in Supporting Information).

Compared with previously identified type-II Dirac materials, the type-II Dirac point in ${\rm ZrO}$ has several advantages.
(i): The Dirac points here almost locate at Fermi energy ($< 5 \ \rm{meV}$), facilitating the experimental observation.
(ii): The linear energy range of Dirac point along tilt direction is large [see Fig. \ref{fig4}(c)], especially for the valence band ($\sim 0.3 \ \rm{eV}$), offering a good platform for studying the intriguing  phenomena associated with type-II dispersion by transport.
(iii): All the three pairs of type-II Dirac points are in the same Fermi energy as they are related to each other by  symmetries, facilitating the experimental identification   of Dirac nodes.

\section{Discussion and conclusion }

From the character table of $C_{3v}$, $C_{4v}$ and $C_{6v}$ point group (without SOC), one knows there  exist both one- and two-dimensional IRs and all the doubly degenerate bands exhibit opposite mirror eigenvalues.
Thus,  according to the above discussion in ${\rm ZrO}$, one can expect that TNP  and TNR  may widely coexist  in the materials with these symmetries.

When SOC is included, the appearance of TNP and TNR requires the absence of $\cal{P}$ symmetry (assuming $\cal{T}$ symmetry maintains).
In the character table of $C_{4v}$ and $C_{6v}$ double group, only two-dimensional IRs exist. Thus,  TNP can not be realized in the rotation axe with $C_{4v}$ and $C_{6v}$ symmetries.
In contrast,  for rotation axe exhibiting  $C_{3v}$ symmetry there still  coexist  one- and two-dimensional IRs and the doubly degenerate  band features opposite mirror eigenvalues.
Thus, in a spinful system with $C_{3v}$ symmetry, the TNP may be realized in the rotation axe and the TNR  may be found in the mirror plane, provided $\cal{P}$ is broken.

Intriguing  magneto-transport phenomena have been studied for a variety of nontrivial band crossings and there exists important distinctions between them.
First we compare the magneto-transport in type-I and type-II TNP. Due to the energy tilt, the Landau level (LL) spectrum of type-II TNP semimetal  is sensitive to  the angle between the magnetic  field and the energy tilt direction. Particularly, when the angle  beyond a critical value, the LL spectrum collapses \cite{Yu2016}.
Such collapse is a unique feature of type-II band crossing and cannot be found in   type-I TNP semimetal.
Then comparing  type-II TNP and type-II Weyl (Dirac) semimetal, we note that the type-II Weyl (Dirac) point only connects one (two identical) electron pocket(s) and one (two identical) hole pocket(s) \cite{Soluyanov2015,Zhou_2016Natphy_MoTe2,Beenakker2016}, whereas the type-II TNP would connect three pockets: either two electron  pockets  and one hole pocket or one electron  pocket  and two hole pockets, as a TNP is formed by three bands.
For example, the TNP shown in  Fig. \ref{fig2}(a) connects two electron  pockets  and one hole pocket.
Because the magnetoresponse near type-II Weyl (Dirac) point and TNP are dominated by the magnetic tunneling between the electron and hole pockets \cite{Beenakker2016}, one can expect that the  magnetic quantum oscillations of   type-II TNP and type-II Weyl (Dirac) semimetal near nodal point would be very different.

Experimentally, the band features predicted here can be detected by the ARPES technique and transport measurements. Due to the strong  metallization of  ZrO, the topological surface states are  deeply buried in the bulk bands and hence are hard to be detected (see Figure S3 in Supporting Information).
However, the bulk type-II Dirac points should be directly observed by ARPES.
In addition, since the type-II Dirac node locates around Fermi energy, it should be identified  by magneto-transport experiment, as its  magnetoresponse is distinguished from type-I semimetal by LL collapses \cite{Yu2016}, from type-II Weyl semimetal by anomalous chiral LLs\cite{2017VAl3} and from type-II TNP semimeatal by magnetic Klein tunneling \cite{Beenakker2016} as discussed above.

In conclusion, we have reported that ${\rm ZrO}$ is a novel topological
metals. When SOC is ignored, three pairs of type-II TNPs and TNRs
coexist in ${\rm ZrO}$. The TNR is formed by three bands and its
evolution against symmetry-preserving strain distinct from that in  two-band NR. Under SOC, each type-II TNP transforms into a Dirac point. Particularly, all the Dirac points reside at same energy and exhibit large linear energy range along their  tilt directions, offering facility for  detecting the interesting properties induced by type-II  dispersion.

\begin{acknowledgements}
The work is  supported by  the National Key R$\&$D Program of China (Grant No.  2016YFA0300600), the MOST Project of China (Grants No. 2014CB920903) and the NSF of China (Grants Nos. 11734003, 11574029).The National Key R$\&$D program under Grant No. 2016YFA0300904, the National Science Foundation of China under Grant No. 61325021, the Key Research Program of Frontier Sciences, CAS under Grant No. QYZDB-SSW-SLH004, and the Strategic Priority Research Program (B), CAS under Grant No. XDPB0602.
\end{acknowledgements}

%
%
%

\bibliography{ZrOref}

\end{document}